
\magnification \magstep1
\hsize=16truecm \vsize=24truecm
\overfullrule 0pt
\parskip=10pt plus 2pt
\fontdimen16\tensy=2.7pt
\fontdimen17\tensy=2.7pt
\rm

\def\picture #1 by #2 (#3){
  \vbox to #2{
    \hrule width #1 height 0pt depth 0pt
    \vfill
    \special{picture #3} 
    }
  }

\def\scaledpicture #1 by #2 (#3 scaled #4){{
  \dimen0=#1 \dimen1=#2
  \divide\dimen0 by 1000 \multiply\dimen0 by #4
  \divide\dimen1 by 1000 \multiply\dimen1 by #4
  \picture \dimen0 by \dimen1 (#3 scaled #4)}
  }

\def\square{\sqcap  \kern-0.69em\hbox{$\sqcup $}}
\def\ssquare{\scriptstyle\sqcap  \kern-0.60em\hbox{$\scriptstyle$}}

\def\hbar{\;{h\mkern-20mu\phantom{o}_{\phantom{1}}^-}\;}
\def\kulma{<\kern-0.8em{\hbox{)}}}
\tolerance=2000
\def\IR{I\kern-0.16em{\hbox{R}}}


\vskip 0.5truecm
\centerline{\bf RENORMALIZATION IN QUANTUM STATISTICS OF SYSTEMS}
\centerline{\bf WITH SPONTANEOUS SYMMETRY BREAKING}
\vskip 0.5truecm
\centerline{M. Chaichian$^a$, J.L. Lucio M.$^b$,
C. Montonen$^c$,}
\centerline{H. Perez Rojas$^{d,}$\footnote{$^1$}{Permanent address: Grupo de
Fisica
Teorica ICIMAF, Calle E  309,  esq. a 15, Vedado, La Habana 4, Cuba.}, M.
Vargas$^b$}

\vskip 1cm
\noindent
a) High Energy Physics Laboratory, Department of Physics
and Research Institute for High Energy Physics,
P.O.Box 9, FIN-00014 University of Helsinki, Finland
\hfill\break
b) Instituto de F'sica de la Universidad de Guanajuato, A.P. E-143, 37150 Leon,
Guanajuato,
Mexico
\hfill\break
c) Department of Theoretical Physics, P.O. Box 9, FIN-00014 University of
Helsinki, Finland
\hfill\break
d) Centro de Investigaci—n y Estudios Avanzados del I.P.N., Departamento de
F'sica,
Apartado 14 - 740, Mexico 07000, D.F.

\vskip 1cm
\hfill\break

\centerline{ABSTRACT}

Renormalization in quantum statistics in the presence of a charge associated to
a
spontaneously broken symmetry is discussed for the scalar field model.  In
contrast
with the case of non-broken symmetry, the renormalization mass counter term
$\delta m^2$
depends on the chemical potential.  We argue that this is connected to the
ill-defined
character of the charge operator.
\hfill\eject
In Ref. [1] it was shown that it is not justified to introduce a charge $Q^N$
associated to a
spontaneously broken symmetry into the density matrix of a grand canonical
ensemble,
because such a charge cannot be considered a constant of motion.  Even in the
classical case
$Q^N$ need not be conserved [2].  Here we consider the renormalization of the
simple
scalar field model in the broken and unbroken phases.  We will argue that the
renormalization is the same in both cases, except for the fact that the mass
renormalization is $\mu$-dependent in the broken phase, ($\mu$ is the chemical
potential connected with $Q^N)$.

Although this result is expected from Symanzik«s theorem [3], we will discuss
it in
some detail, as in the recent literature [4] there are claims in the opposite
sense.  It is
important to note that such a $\mu$-dependent countertem makes the average
charge
$< Q^N >$ formally infinite in the broken phase, reflecting that $Q^N$ is
ill-defined in
the broken phase [5].

After functional
integration over canonical momenta the system is described by the partition
function

$$Z = N(\beta)\int D\phi^+ D\phi^- \exp \int dx_4 \int d^3 x {\cal L}_{eff}
\eqno(1a)$$
where $N(\beta )$ is an unimportant constant and ${\cal L}_{eff}$ is the
effective
Lagrangian
$${\cal L}_{eff} = -(\partial_\nu - \mu \delta_{\nu 4}) \phi^- (\partial_\nu +
\mu\delta_{\nu 4}) \phi^+ - m^2 \phi^- \phi^+ - \lambda^2 (\phi^- \phi^+ )^2
\eqno (1b)$$
where $\phi^\pm = (\phi_1\pm i\phi_2 )/\sqrt 2.$\hfil\break

If there is no symmetry breaking, i.e. $m^2 > 0$ and $<\phi^\pm > = 0 ,$ the
divergences of the
theory are contained in terms independent of $\mu$ and $T$, in other words, the
renormalization constants $Z_1 , Z_2$ and $\delta m^2$ are the same as those
obtained
in the Euclidean quantum field theory corresponding to the zero temperature,
zero chemical
potential limit.

Consider for example the tadpole term in the perturbative expansion of the
temperature
Green function
$$G_{il}(x,x') = G_{0il}(x,x') - {1\over 2}\lambda^2 \int d^4 y  G_{0ij}(x_1
-y)
G_{0jk}(y-y)G_{0kl}(y-x')+\cdots \eqno(2)$$
In the diagonal representation of $G_{0kl}$ we have

$$G_0(0) = \lambda^2(2\pi^2 )^{-1} \sum_{k_4} \int k^2 dk G^\pm (k),  \eqno
(3a)$$
where

$$G^\pm (k) = \left( \matrix{((k_4 -i\mu )^2 + {{\cal E}_k}^2 )^{-1} & 0 \cr
0 & ((k_4 + i\mu )^2 + {{\cal E}_k}^2)^{-1}\cr} \right)$$
and therefore

$$G_0 (0) = -\lambda^2(2\pi^2)^{-1} \int k^2 {dk{\cal E}_k}^{-1} (n^+ + n^- -1)
\eqno (3b)$$
with ${\cal E}_k = (k^2 + m^2 )^{1\over 2}$ and $n^\pm = (\exp ({\cal E}_k \mp
\mu )
\beta - 1)^{-1}. $ Ultraviolet divergences arise only in the $\mu , \beta$
independent term.
This means that  renormalization can be achieved by introducing $\mu , \beta$
independent
counterterms.

It should be noted that eq. (3b)  (and eq. (6) below) restrict the allowed
range of
$\mu$ [6] even in the interacting theory as otherwise the integrands develop
poles
through zeros in $\varepsilon_k \pm \mu$ or $\varepsilon_\pm$.  This makes the
discussion concerning possible phase transition for $\mu^2 >> m^2 $ [4, 7]
irrelevant.

The same diagram calculated after the symmetry breaking, i.e. for
$m^2 = -a^2 ,$  leads to an expression analoguous to (3a) in some diagonal
representation,
but now

$$G^\pm (k) = \left( \matrix{(k^2_4 + {\cal E}^2_+ )^{-1} & 0 \cr
0 & (k^2_4 + {\cal E}^2_- )^{-1}\cr} \right) \eqno (4)$$
where
$${\cal E}_\pm^2 = \{(E^2_1 + E^2_2 + 4\mu^2 ) \pm [(E^2_1 + E^2_2 + 4\mu^2 )^2
-4 E_1^2 E_2^2 ]^{1\over 2} \}/2  \eqno (4a)$$
and at the tree level
$E^2_1 = k^2 + 2(a^2 + \mu^2 ), E_2^2 = k^2 .$

Let us consider the terms in the diagonal of (4).  After summation over $k_4$
they
lead respectively to
$$[n({\cal E}_\pm ) + {1\over 2}]/{\cal E}_\pm ,$$
where

$$n({\cal E}_\pm ) = [\exp ({\cal E}_\pm \beta ) - 1 ]^{-1} \eqno (5)$$
Notice that in the zero temperature limit the diagonal terms lead also to
divergences,
however in this case they depend through ${\cal E}_\pm$ on the chemical
potential.
We have in the diagonal representation in which we are working

$$G_0 (0)_{11,22} =  -\lambda^2 (2\pi^2 )^{-1} \int k^2 dk {\cal E}_\pm^{-1} ,
\
G_{0ij} = 0 \ {\rm if} \  i \not= j  \eqno (6)$$
The leading terms in the large momentum expansion of ${\cal E}_\pm^{-1}$ are
given by

$${\cal E}_\pm^{-1} = {2\over k} \left[1-{a^2 + 3\mu^2\over 2k^2} \mp
{2\mu\over k} \left( 1+ {a^2 + 3\mu^2\over 8\mu k^2}\right)\right] \eqno (7)$$

We conclude that eq. (6) contain $\mu$ dependent divergences and in
consequence,
$\mu$ dependent counterterms must be added to the Lagrangian (1b).

Moreover, if we consider the zero temperature limit of the thermodynamic
potential
we have

$$\Omega = (4\pi^2 )^{-1} \int k^2 dk ({\cal E}_+ + {\cal E}_- ) \eqno (8)$$
where

$${\cal E}_+ + {\cal E}_- = \sqrt 2 \{k^2 + a^2 + 3\mu^2 + k[k^2 + 2(a^2 +
\mu^2 )]^
{1\over 2} \}^{1\over 2} \eqno (9)$$
Expanding in powers of $k^{-1}$ it is easily seen that $\Omega $ contains also
$\mu$
dependent divergences. \hfil\break
A different approach to this problem was adopted by Benson, Bernstein and
Dodelson in [4],
where  they conclude that (8) contains only $\mu$  independent divergences.
Their argument
is  based on dimensional analysis for which they take the expectation value
$\xi^2 = <\phi^2 >$ as well as $\mu^2 , a^2 , \lambda$ as {\it independent}
parameters.
But that amounts in effect to working away from the minimum of the effective
potential, i.e.
taking the particles off their mass shell.  We will argue below that a proper
calculation of
the thermodynamic potential $\Omega$ requires working at the minimum of the
effective
potential where the vacuum expectation value $\xi$ is given in terms of $\mu$,
$a$ and
$\lambda$.  In fact, the relation between the parameters is a fundamental
consequence of
the spontaneous symmetry breaking and it also implies the existence of a
Goldstone boson.
We will show, following Fradkin and Tyutin [8], that if we consider the broken
case of the
$\mu$-dependent Euclidean quantum field theory defined by the Lagrangian (1b),
then: i)  As in the $\mu = 0$ case, the Goldstone theorem results as a
consequence of the
Ward identities,  ii) The exact value of $\xi$ is given as a function of the
parameters of
the theory $\lambda^2 , m^2 = -a^2 , \mu .$

By expressing (1b) in terms of the fields $\phi_1 ,\phi_2$ and introducing the
external
currents $J_1 , J_2 ,$ we may write the generating functional

$$W(J_i ) = -\beta^{-1} ln Z(J_i ) \eqno (10)$$
where

$$Z(J_i ) = N(\beta ) \int D\phi_1 D\phi_2 \exp \int dx_4 \int d^3 x ({\cal
L}_{eff} +
J_i \phi_i ) \eqno (11) $$
The average fields are $\varphi_i = <\phi (J_i )> = \delta ln W/\delta J_i $
and the
Green functions $G_{ij} = \delta \varphi_i / \delta J_j = \delta^2 ln W /
\delta J_i
(x)\delta J_j (y). $  Based on the global $U(1)$ invariance of the model we get
the Ward
identities

$$\int d^4 x J_i (x) \varepsilon_{ij} \varphi_j (x) = 0 \eqno (12)$$
where $\varepsilon_{12} = -\varepsilon_{21} = -1;\  \varepsilon_{11} =
\varepsilon_{22} =
0.$ From (12) by differentiating functionally with respect to the fields
$\varphi_i$ we
get $(\varphi_1 = \xi , \ \varphi_2 = 0$ for $J_i = 0):$

$$\eqalign {\xi G_{22}^{-1}  (p = 0, \xi ) = & \xi G_{21}^{-1} (p = 0, \xi )  =
\xi G_{12}^{-1} (p = 0 , \xi ) = 0\cr
G_{11}^{-1} (p = 0,\xi ) = & \partial [\xi G_{22}^{-1} (p = 0,\xi )]
/\partial \xi \cr} \eqno (13)$$
The same set of equations can be obtained by starting from the renormalized
Lagrangian.  The set (13) then shows that i)  the mass matrix has a zero
eigenvalue and
ii) $\xi$ is given exactly as a function of the parameters of the theory
$a^2, \lambda^2 ,\mu .$ (at the tree level $\xi^2 = (a^2 + \mu^2 )/\lambda^2$
arises from
$G_{22}^{-1}(k^2 ) = k^2 + (a^2 + \mu^2 - \lambda^2 \xi^2 )$ by writing
$G_{22}^{-1} (0) = 0).$

We have explicitly checked the arguments presented so far, by performing the
one-loop
renormalization of the model both in the broken and unbroken phases.  The
analysis of the
unbroken phase is straightforward, the result coinciding with the $\mu = 0$
case where
the renormalization constants are given by:

$$Z_\phi = 1,\ Z_m = 1 + 4\lambda \Gamma (\varepsilon ), \ Z_\lambda = 1 + 10
\lambda \Gamma (\varepsilon ) \eqno (14)$$
where $\Gamma (\varepsilon ) = 1/\varepsilon + 0(\varepsilon^0 ),$ and to get
(14) we
used dimensional regularization.

In the broken phase we need to consider only the mass renormalization since the
one
loop $\mu$-dependent corrections to the field $\phi$ and coupling constant
$\lambda$ are finite.  In the $\phi '_1 ,\varphi_2 $ basis the propagators are
obtained from

$$(\phi'_1 ,\phi_2 ) \left( \matrix{k^2_4 + k^2 + 2\lambda\xi^2 & -\mu k_4\cr
\mu k_4 & k_4^2 + k^2\cr}\right)
\left(\matrix{\phi_1'\cr \phi_2\cr}\right) \eqno (15)$$
The one-loop corrections to the diagonal terms of the propagators resulting
from (15)
are readily calculated $(m_1^2 = 2 \lambda \xi^2 ,$ since we are working at the
tree mass shell):

$$Z_1 = Z_2 = 1; \quad Z_{m_1} = 1 + 4 \lambda \Gamma (\varepsilon ) -
8\lambda (\mu^2 /2\xi^2 ) \Gamma (\varepsilon ) \eqno (16)$$
while the Goldstone boson remains massless, as the Ward identities (13) demand.
 Notice
that (16) agrees with (14) only if $\mu = 0.$  We could argue that $\mu$ should
be
renormalized, however, explicit calculations of the corrections of the
off-diagonal
terms to (15) shows that they vanish.

Thus this theory requires the same type of counterterms as the
theory without chemical potential, however, we see that one of these
counterterms is
$\mu$-dependent.  In other words, the model is renormalizable but at the price
of
introducing some $\mu$-dependent counterterm.  This means that
renormalization prescriptions demand the introduction of some additional $\mu$-
dependent term in the exponent of the density matrix from which we started,
i.e. we
must write ${\cal H}-\mu Q^N -\delta m^2 (\mu , a^2 ,\lambda^2 )\phi^+ \phi^- +
\dots$  But this implies that we are renormalizing the vacuum charge density
$<Q^N >$ by subtracting from it some divergent term, the "charge" density
$\partial < \delta m^2\phi^+ \phi^- +\dots >/\partial \mu$.

The ${\mu}$-dependent divergence is very similar to the one arising in the
Euler-
Heisenberg vacuum term in electrodynamics.  In that case, divergences are
present in
the initial energy expression as a consequence of the vacuum fluctuations of
the
external field, and must be removed by subtracting divergent terms depending on
the
external electromagnetic field.  Our case has the additional ingredient that
the
"external field" $<\phi^2 > = \xi^2$ depends on the chemical potential $\mu ,$
and the
subtraction procedure must involve a $\mu$-dependent term.

Concerning the very interesting result obtained in Ref. [4] on the
non-relativistic
limit of the spectrum of the model described by (1) in the SSB case, leading to
the
hard-sphere boson gas spectrum, it is clear that our expression (4a) leads to
the same
non-relativistic limit.

The difference in the behaviour of the broken and unbroken phases can be partly
understood as follows:  In the symmetric case the vacuum state $|0>$ is unique,
and a
Hilbert space of states containing states of definite charge can be built on
it.  The charge
operator $Q$ is well-defined in this space, (more exactly the operator $Q_V$
measuring
the charge in a finite volume $V$ has a well defined limit $\rho = \lim_{V\to
\infty} Q_V)$
and annihilates the vacuum.  The statistical average is a weighted average over
this space
of states.

In the broken phase the situation is radically different [9,1,2].
We now have an infinity of possible vacua, and on each we can build a space of
states
carrying a representation of the operator algebra, not equivalent to any
representation
built on a different vacuum.  The charge operator, which formally generates an
intertwining operator between these inequivalent representations $(Q^N|{\rm
vac} >
\neq 0)$ cannot be represented in any of the state spaces, and becomes
meaningless.
Our results indicate that in the perturbative loop expansion considered here,
this is
reflected in an infinite value of the average charge $< Q^N >.$  We expect that
this state of
affairs remains true for gauge theories as well, implying e.g. that it is
meaningless to
introduce a chemical potential coupled to the weak neutral charge in the
standard model.

The authors thank J. Bernstein for correspondence and
several illuminating remarks, A. Cabo and A. Gonz‡lez for fruitful
discussions, and A. Zepeda for comments.  The authors J.L.L.M., M.V. thank
ICIMAF and C.M. thank
ICIMAF and ISPJAE for hospitality in Havana, whereas H.P.R. thanks CONACyT for
financial support and CINVESTAV for hospitality.  J.L.L.M and M.V. were
supported by CONACyT
under contracts  F246 - E9207 and 1628 - E9209.
\bigskip
References:
\item {1)} M. Chaichian, C. Montonen and H. Perez Rojas, Phys. Lett. B 256
(1991) 227.
\item {2)} M. Chaichian, J.A. Gonz‡lez, C. Montonen and H. Perez Rojas, Phys.
Lett. B 300
(1993) 118.
\item {3)} K. Symanzik, in Cargse Lectures in Physics, vol. 5, D. Bessis ed.
(Gordon and
Breach, New York 1971).
\item {4)} K.M. Benson, J. Bernstein and S. Dodelson, Phys. Rev. D 44 (1991)
2480;
J.  Bernstein and S. Dodelson, Phys. Rev. Lett. 66 (1991) 683.
\item {5)} L.P. Horowitz and S. Raby, Phys. Rev. D 15 (1977) 1772; see also
F. Strocchi, Elements of Quantum Mechanics of Infinite Systems (World
Scientific,
Singapore, 1985).
\item {6)}  H.E. Haber and H.A. Weldon, Phys. Rev. Lett. 46 (1981) 1947; Phys.
Rev.
D25 (1982) 502.
\item {7)} S. Mohan, Phys. Lett. B 307 (1993) 367.
 \item {8)} E.S. Fradkin, I.V. Tyutin, Riv. del Nuovo Cim. 4 (1974) 1.
\item {9)} R. Haag, Nuovo Cimento 19 (1962) 287.

\end